\documentclass[slac_one]{revtex4}
\usepackage[dvips]{graphicx}
\usepackage{fancyhdr}
\begin{document}

%Title of paper
\title{{\small{Hadron Collider Physics Symposium (HCP2008),
Galena, Illinois, USA}}\\ %% Please keep this conference title here
\vspace{12pt}
Other BSM Searches at the LHC} %% Paper title goes here

% Repeat the \author .. \affiliation  etc. as needed
%
% \affiliation command applies to all authors since the last
% \affiliation command. The \affiliation command should follow the
% other information

\author{D.Bourilkov\\(for the ATLAS and CMS Collaborations)}
\affiliation{University of Florida, P.O. Box 118440,
Gainesville, FL 32611, USA}

\begin{abstract}
A review of the discovery potential of LHC for new phenomena beyond
the Standard Model (BSM) other than Supersymmetry in the early phase of
running is presented. Topics covered include searches for extra
dimensions in different scenarios (ADD, Randall-Sundrum,
black holes ...), resonance hunting in di-lepton,
di-photon and di-jet final states, searches for contact interactions,
heavy stable charged particles, technicolor, etc. The strategies of
the ATLAS and CMS experiments to understand the detectors 
and prepare them for "search" mode and the prospects for
discoveries using early data are described.
\end{abstract}

%\maketitle must follow title, authors, abstract
\maketitle

\thispagestyle{fancy}

% body of paper here - Use proper section commands
% References should be done using the \cite, \ref, and \label commands
% Put \label in argument of \section for cross-referencing
%\section{\label{}}

\section{INTRODUCTION} % Section title should be in all capitals.
At the dawn of the LHC era particle physics finds itself in a 
unique and paradoxical situation. On the credit side we have a
Standard Model (SM) which could be the envy of many disciplines - 
tested to tantalizing detail and shown to explain
well the vast expanse of often very precise experimental results,
on the debit side no one seems to be happy with it.
Of course there are weighty arguments like the unification of forces
or the glaring omission of gravity, just to name a few, that point
to new physics at higher energies. Moreover there are strong expectations
that something new will happen at energies around a couple of TeV, the
Terascale that LHC is about to start exploring. These enhanced
expectations have led to the development of a flurry of models,
and the hope is that the LHC will serve as Occam's razor and point
the way beyond the Standard Model.

The scope of this talk is to cover all other BSM searches except the
200 kg gorilla better known as Supersymmetry. This is a huge field, I will
focus on ``clean'', low background channels which can provide early
discoveries once the machine, detectors and Standard Model contributions
are understood in sufficient detail. Advance apologies for the omissions,
unavoidable in a short review. Most results are for 14 TeV center-of-mass
energy, but luminosities $\sim$~40 pb$^{-1}$ at 10 TeV have interesting
potential. And we better be ready from the start: the W and Z were
discovered by UA1/UA2 with $\sim$~20/55 nb$^{-1}$, of course they knew where
to look.

\section{BSM SEARCHES}

The results are organized by experimental signatures, not by the models.
For details the readers are referred to~\cite{ATLASpub,ATLASexo,
CMSPTDR1,CMSPTDR2,CMSexo}.

\subsection{One Lepton}

\begin{figure}[!Hhtb]
  \centering
  \begin{tabular}{cc}
    \resizebox{0.49\textwidth}{8.2cm}{\includegraphics{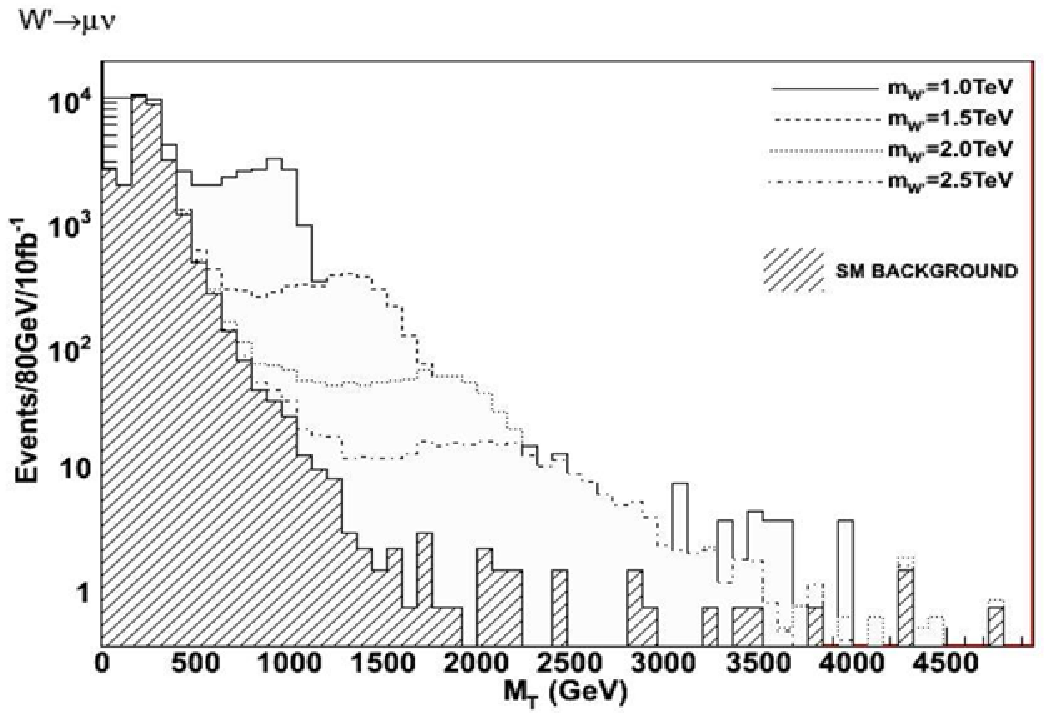}}
    \resizebox{0.49\textwidth}{8.2cm}{\includegraphics{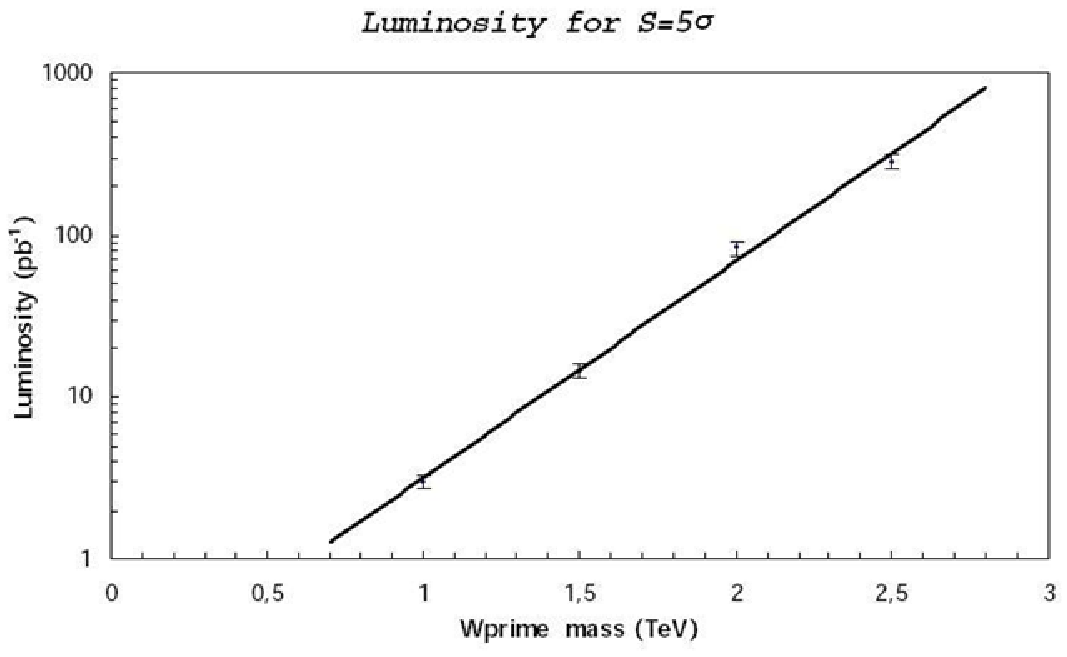}}
  \end{tabular}
\vspace*{-3pt}
\caption{ATLAS W' muon study: Left (1a): M$_T$ distribution.
Right (1b): discovery reach as function of luminosity.}
\label{fig:Fig1}
\end{figure}

The channel with one lepton (electron or muon) and missing transverse energy,
used to discover the W boson, is ideal search field for an additional charged
gauge boson W'. Both collaborations have new results for the muon channel,
showing that very little luminosity is needed in the case of the Sequential
Standard Model (SSM), see e.g. Figure~\ref{fig:Fig1}. While we are unable to
reconstruct the full invariant mass, the transverse mass M$_T$ is a powerful
discriminator. The SM backgrounds are the usual suspects which appear throughout
this talk, I will list them once here and only mention additional background
contributions for specific channels later:
\begin{itemize}
 \item $W \rightarrow \mu\nu_{\mu}$
 \item $Z \rightarrow \mu\mu$
 \item Di-jet QCD
 \item $t\bar{t}$
 \item $WW,\ WZ,\ ZZ$.
\end{itemize}

\subsection{LHC as Bump Hunting Factory}

\begin{figure}[!Hhtb]
  \centering
  \begin{tabular}{cc}
    \resizebox{0.49\textwidth}{8.0cm}{\includegraphics{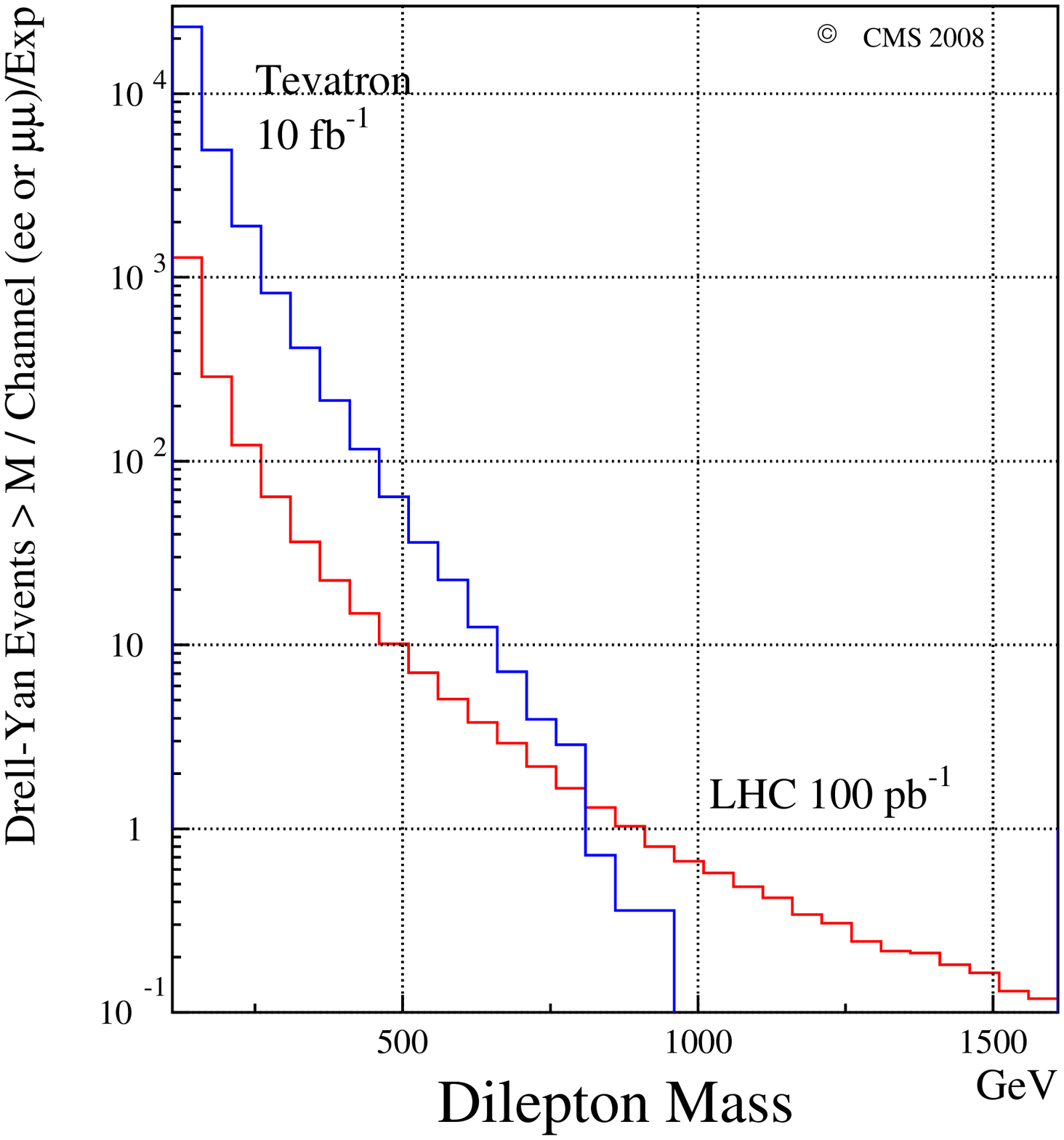}}
    \resizebox{0.49\textwidth}{8.0cm}{\includegraphics{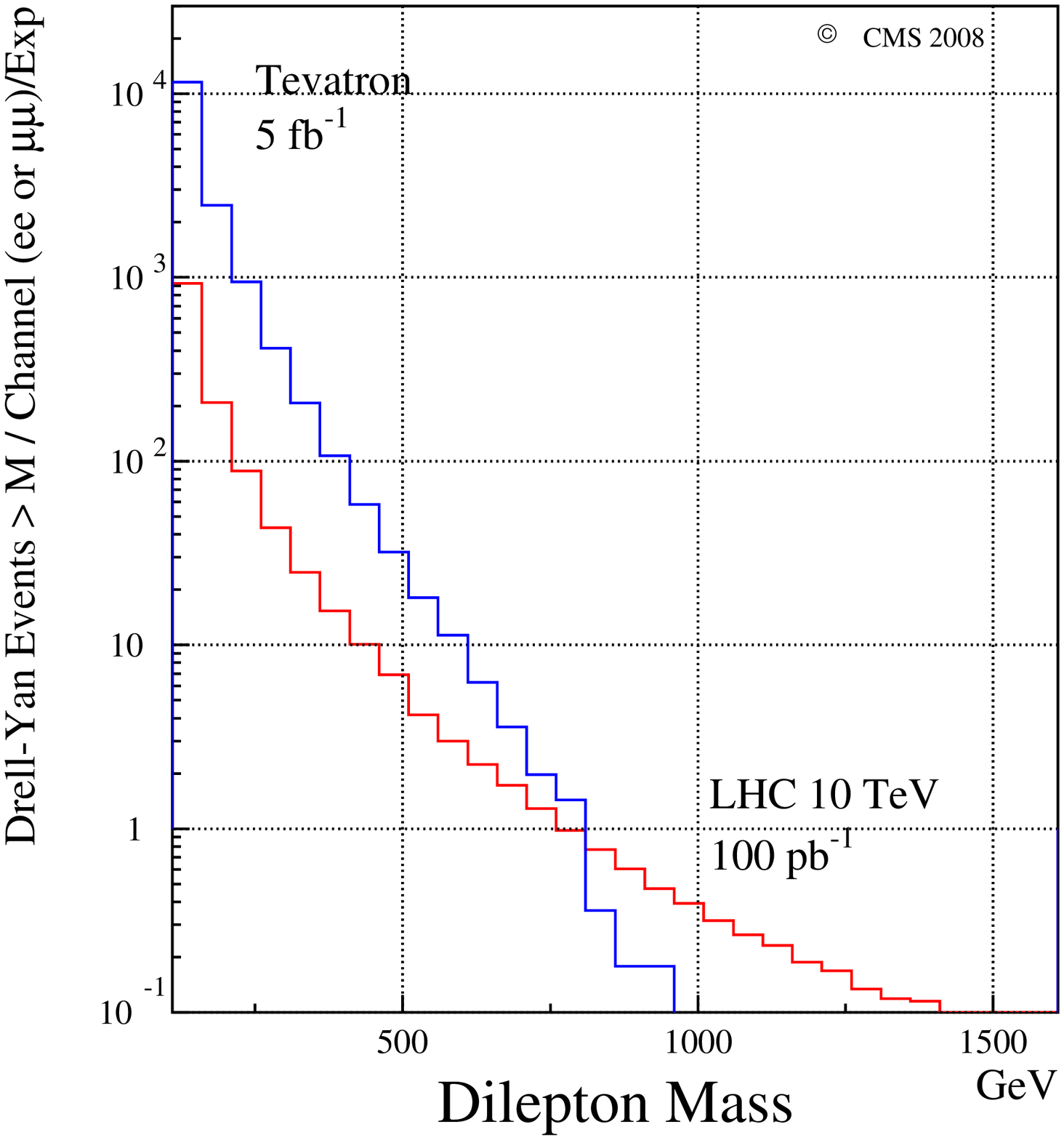}}
  \end{tabular}
\vspace*{-3pt}
\caption{Cumulative plots for the expected number of events at the LHC
{\em above} a given mass in a single Drell-Yan channel (di-electron or
di-muon) for one experiment and comparison with the Tevatron (rough
estimate).
 Left (2a): LHC at nominal energy.
Right (2b): LHC at 10 TeV.}
\label{fig:Fig2}
\end{figure}

A primary early search field are simple final states where complete reconstruction
of the invariant mass of new objects, in the ideal case ``narrow'' bumps, is feasible.
Due to the large parton luminosity up to TeV masses the LHC is a di-\{lepton, photon, jet\}
factory, able to test the SM up to the highest available momentum transfers and search for
signals for many new physics scenarios: ``easy'' like resonances (Z', Randall-Sundrum (RS)
gravitons ...), or ``not-so-easy'': non-resonant deviations from the SM or just tails
like compositeness, extra dimensions in the ADD scenario, etc. In Figure~\ref{fig:Fig2}
rough estimates for the rate of Drell-Yan events is shown, illustrating how fast the LHC
will start probing masses unaccessible at the Tevatron. As a rule of thumb, at 10 TeV
center-of-mass energy
the dominant Drell-Yan background for 10 pb$^{-1}$ is negligible (less than one expected event
per channel) above 500 GeV, for 100 pb$^{-1}$ above 1 TeV. Any statistically
significant accumulation of events in these areas would signal new physics.

\subsection{Two Leptons}

\begin{figure}[!Hhtb]
  \centering
  \begin{tabular}{ccc}
    \resizebox{0.40\textwidth}{7.2cm}{\includegraphics{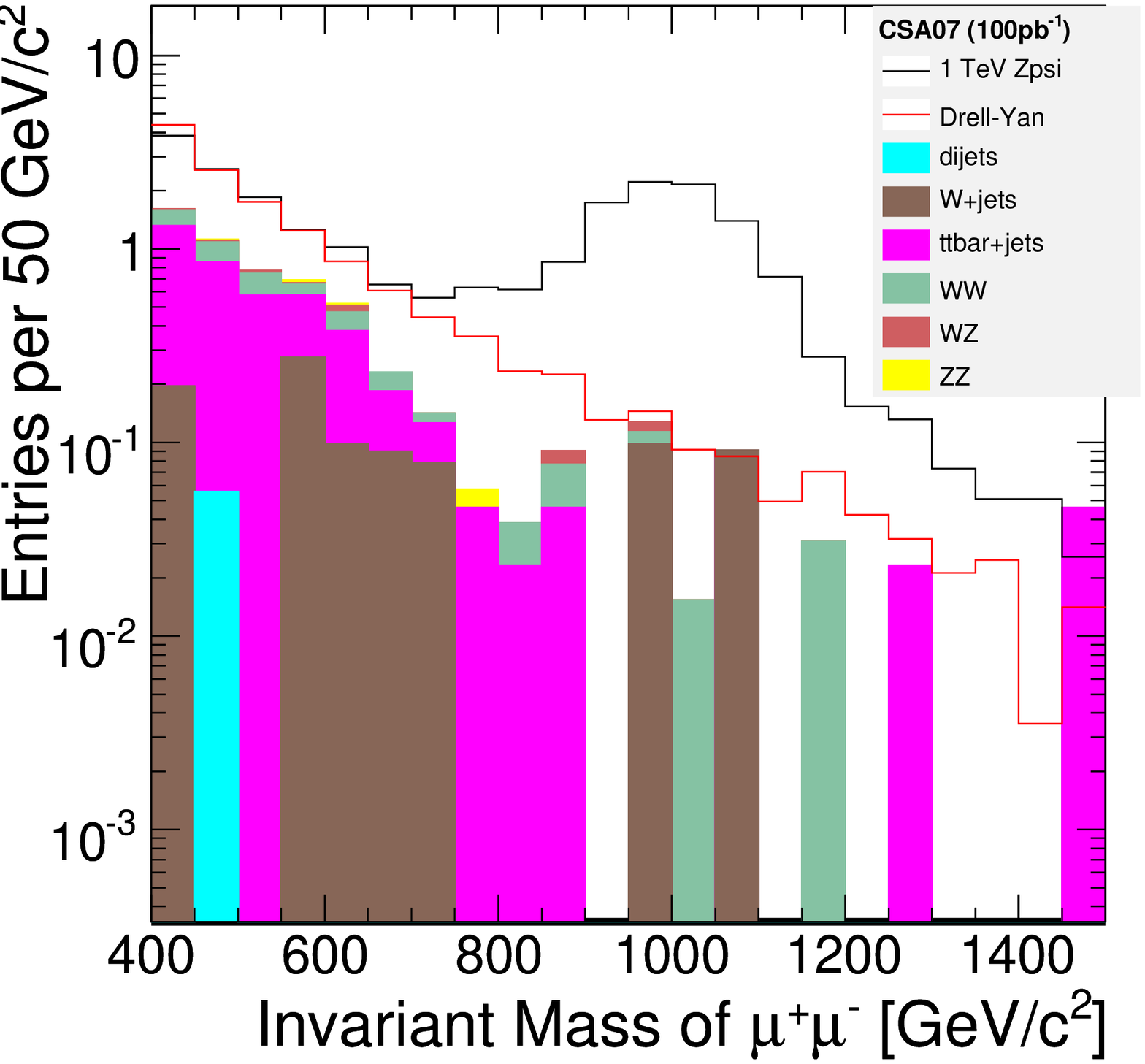}}
    \resizebox{0.29\textwidth}{7.2cm}{\includegraphics{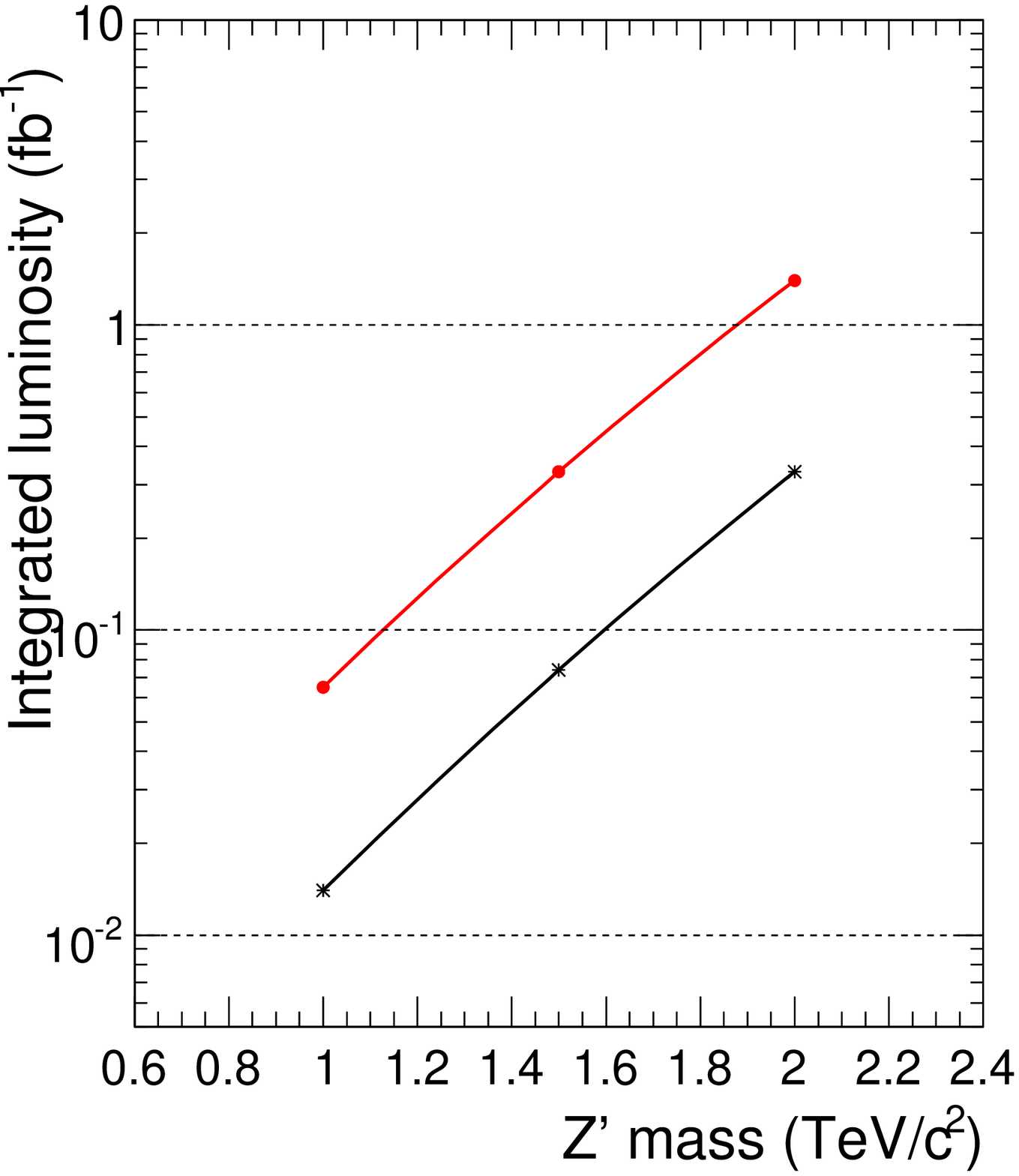}}
    \resizebox{0.29\textwidth}{7.2cm}{\includegraphics{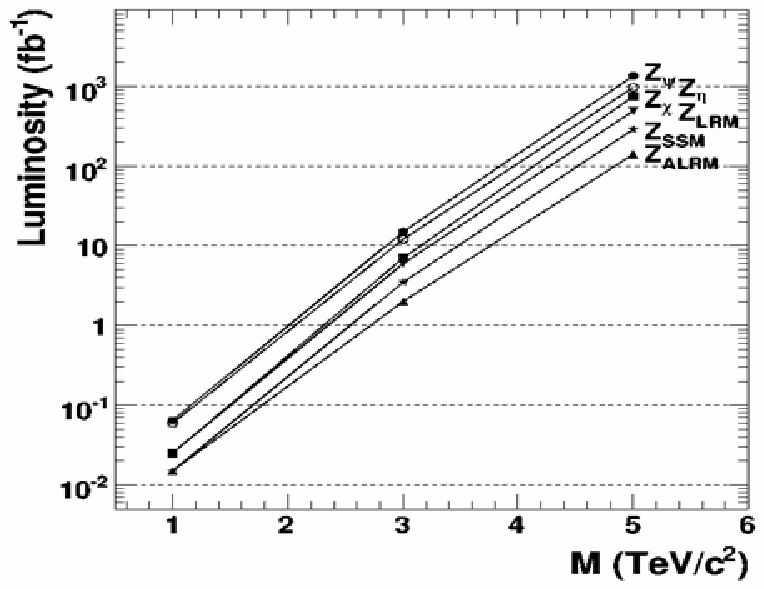}}
  \end{tabular}
\vspace*{-3pt}
\caption{CMS Z' study: Left (3a): backgrounds in the di-muon channel.
Middle (3b): di-muon discovery reach as function of luminosity.
Right (3c): di-electron discovery reach as function of luminosity.}
\label{fig:Fig3}
\end{figure}

The two lepton channel is the classical hunting field where $J/\psi, \Upsilon$ and $Z$
where discovered. An additional gauge boson Z' decaying to di-electrons or di-muons
appears in many scenarios. An example of the various backgrounds and their magnitude
and the discovery reach for the two channels is shown in Figure~\ref{fig:Fig3} for CMS
and different Z' models. Numbers to keep in mind are that LEP2 has already excluded a
SSM Z' below 1.8 TeV and cut deeply into the LHC phase space, and the Tevatron is
approaching the kinematic limit for several models.

\begin{figure*}[t]
\centering
        \includegraphics[width=0.90\textwidth,height=10.0cm]{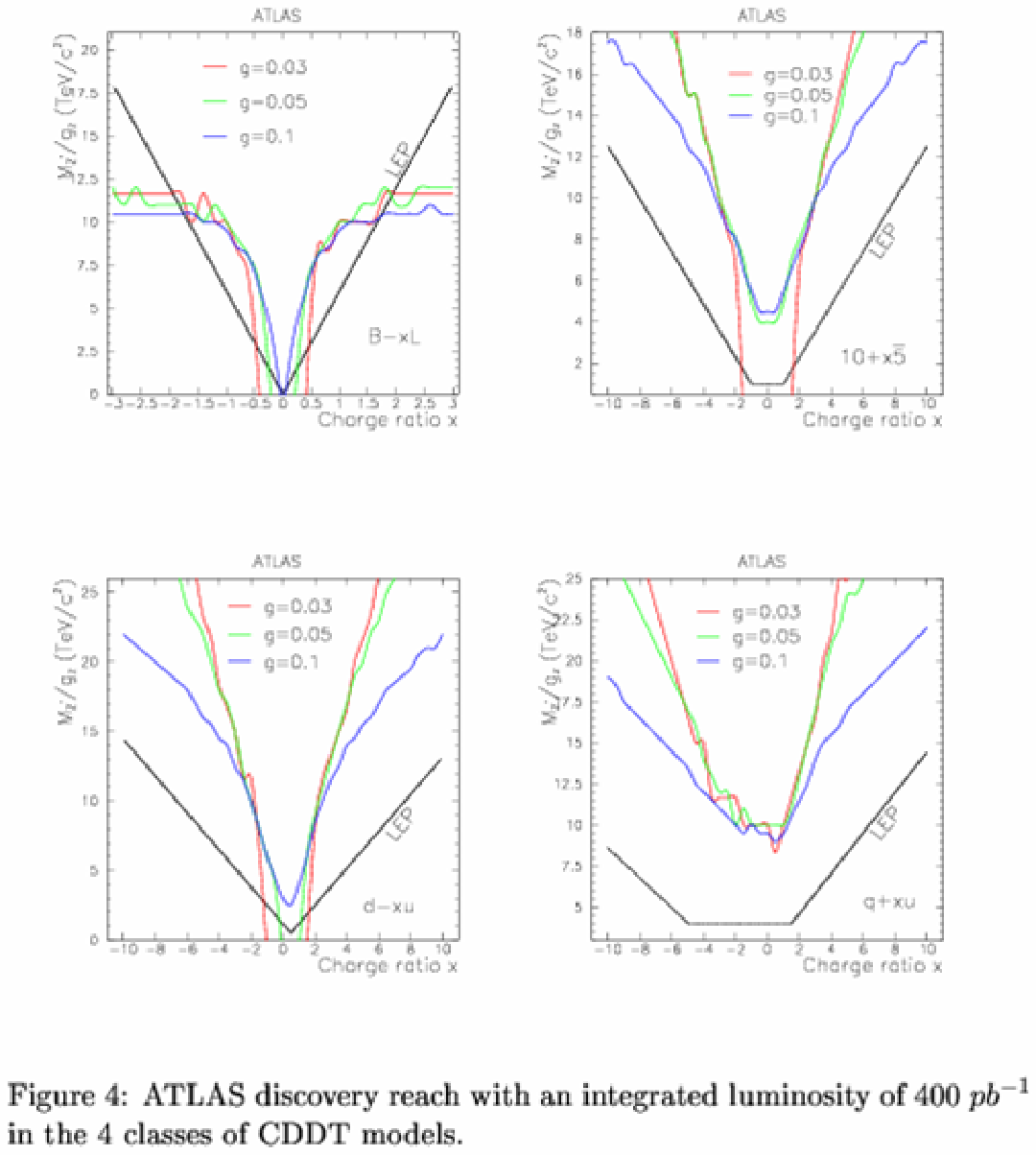}
\caption{ATLAS Z' study.}
\label{fig:Fig4}
\end{figure*}

ATLAS has performed a study using the nearly model-independent CDDT parameterization.
The results, shown in Figure~\ref{fig:Fig4} depend on three parameters:
the mass of the gauge boson $M_{Z'}$, the global coupling strength $g_{Z'}$ and the
relative coupling strength to different fermions $x$. Four classes of solutions are
possible, denoted as B-xL, 10+x\underline{5}, d-xu, q+xu. As can be seen the LEP limits
will be superseded very fast.

\subsection{Two Photons}

\begin{figure}[!Hhtb]
  \centering
  \begin{tabular}{cc}
    \resizebox{0.49\textwidth}{7.6cm}{\includegraphics{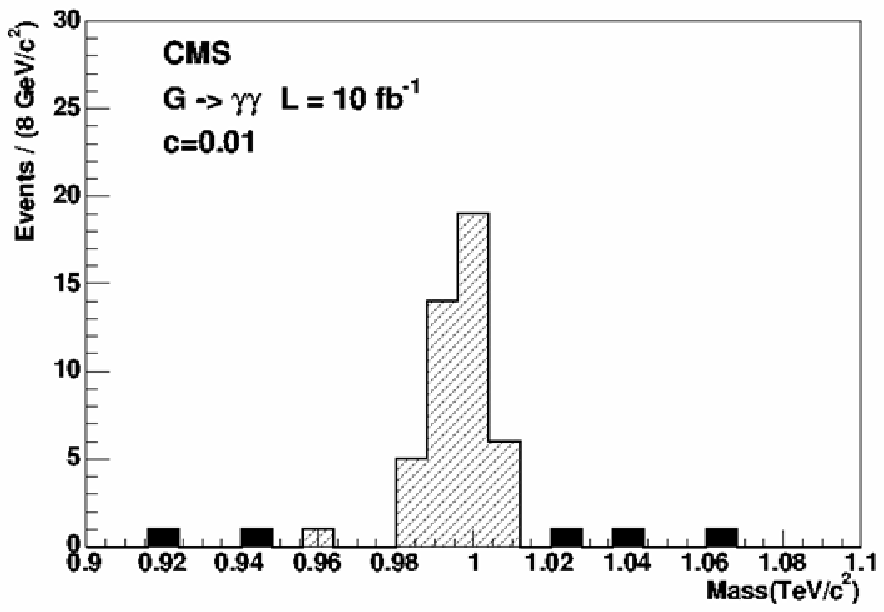}}
    \resizebox{0.49\textwidth}{7.6cm}{\includegraphics{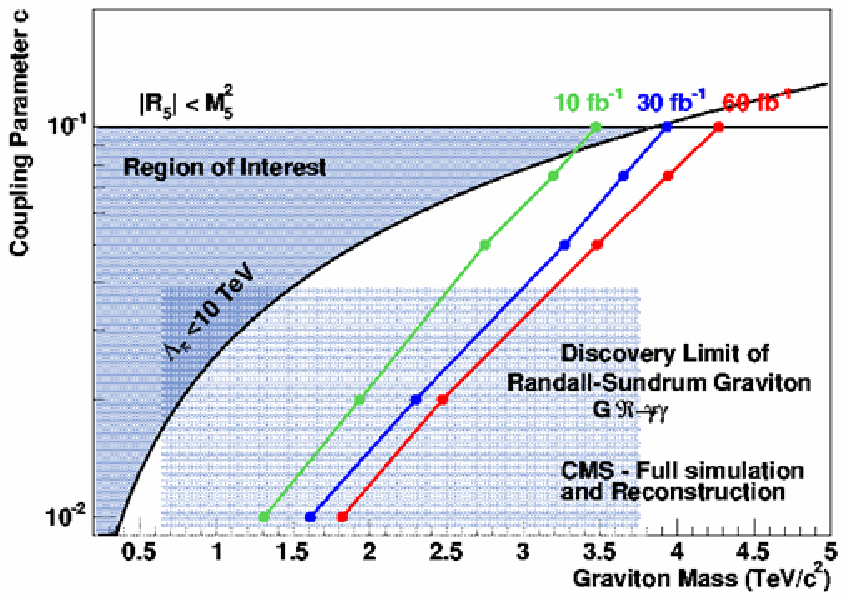}}
  \end{tabular}
\vspace*{-3pt}
\caption{CMS di-photon study: Left (5a): example of invariant mass distribution.
Right (5b): discovery reach for the Randall-Sundrum model as function of luminosity
and coupling strength.}
\label{fig:Fig5}
\end{figure}

The excellent resolution of the detectors makes searches with photons very
competitive, as the physical backgrounds are typically smaller, the
machine backgrounds and fakes need to be well under control, The large
signal to noise ratio and the discovery reach for narrow Randall-Sundrum gravitons
are illustrated in Figure~\ref{fig:Fig5}.

\subsection{Two Jets}

\begin{figure}[!Hhtb]
  \centering
  \begin{tabular}{cc}
    \resizebox{0.49\textwidth}{8.0cm}{\includegraphics{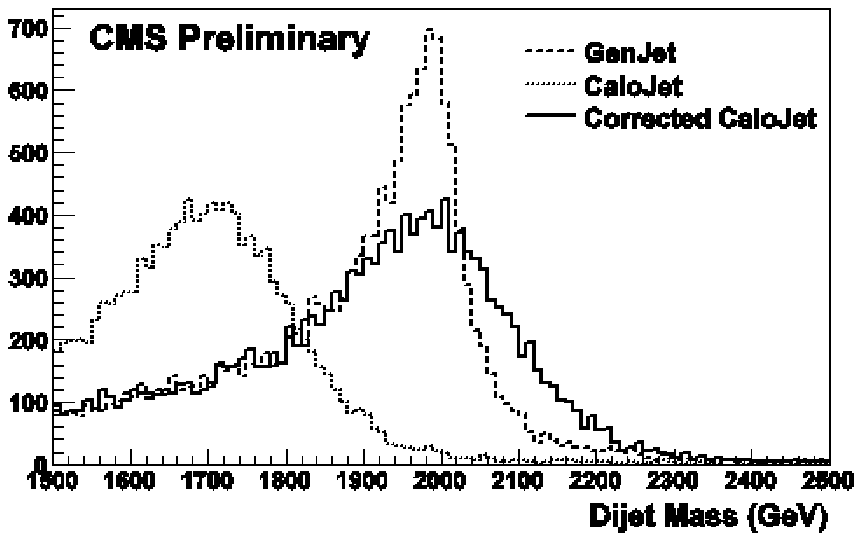}}
    \resizebox{0.49\textwidth}{8.0cm}{\includegraphics{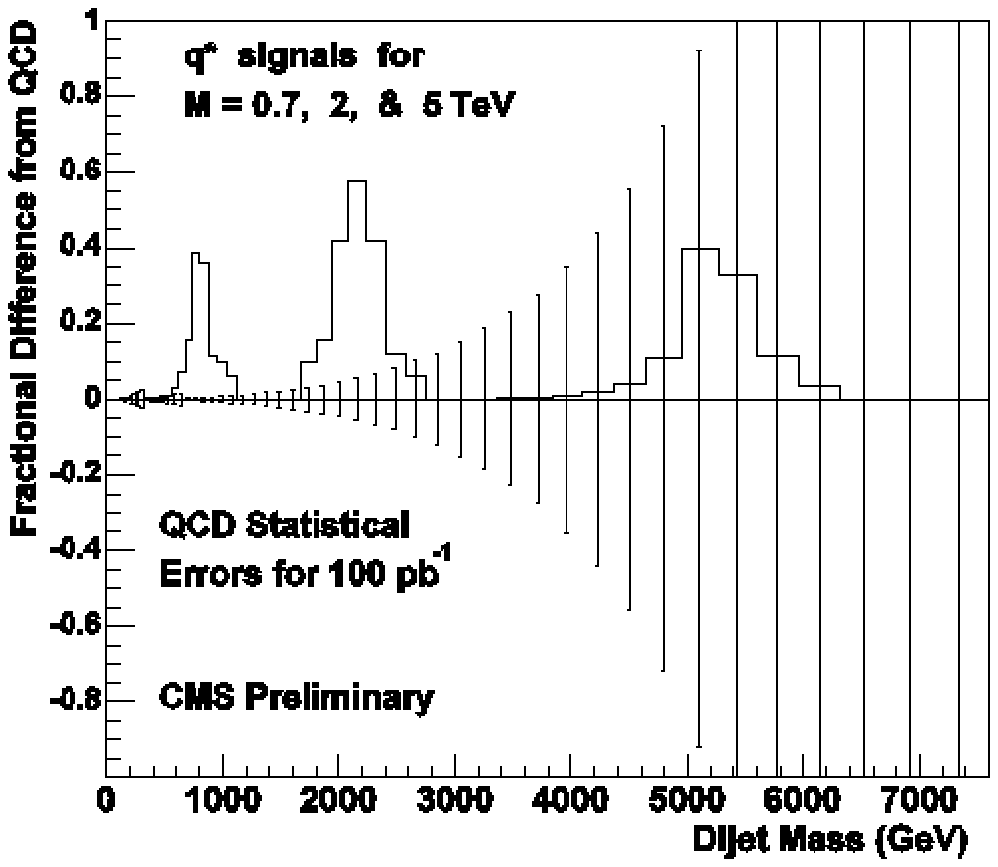}}
  \end{tabular}
\vspace*{-3pt}
\caption{CMS di-jet study: Left (6a): example of invariant mass distribution before
and after corrections.
Right (6b): discovery reach for excited quarks and 100 pb$^{-1}$.}
\label{fig:Fig6}
\end{figure}

A more difficult channel with large discovery potential, the di-jet final state
is sensitive to the jet energy scale and large QCD background, typical for a
hadron collider. The parton density functions (PDF) have to be understood
well to avoid being fooled by initial state effects. The invariant mass
reconstruction and the early discovery potential are shown e.g. in
Figure~\ref{fig:Fig6} for CMS.

\subsection{Three Leptons}

This is a more tricky channel. It can be used to search e.g for technicolor models
where the ``walking'' gauge coupling lowers the scale, and light almost degenerate
vector techimesons can be observed. They decay 
\begin{equation}\label{eq:units}
\rho_T/a_T \rightarrow WZ \rightarrow 3l + \nu.
\end{equation}

\begin{figure}[!Hhtb]
  \centering
  \begin{tabular}{cc}
    \resizebox{0.49\textwidth}{8.90cm}{\includegraphics{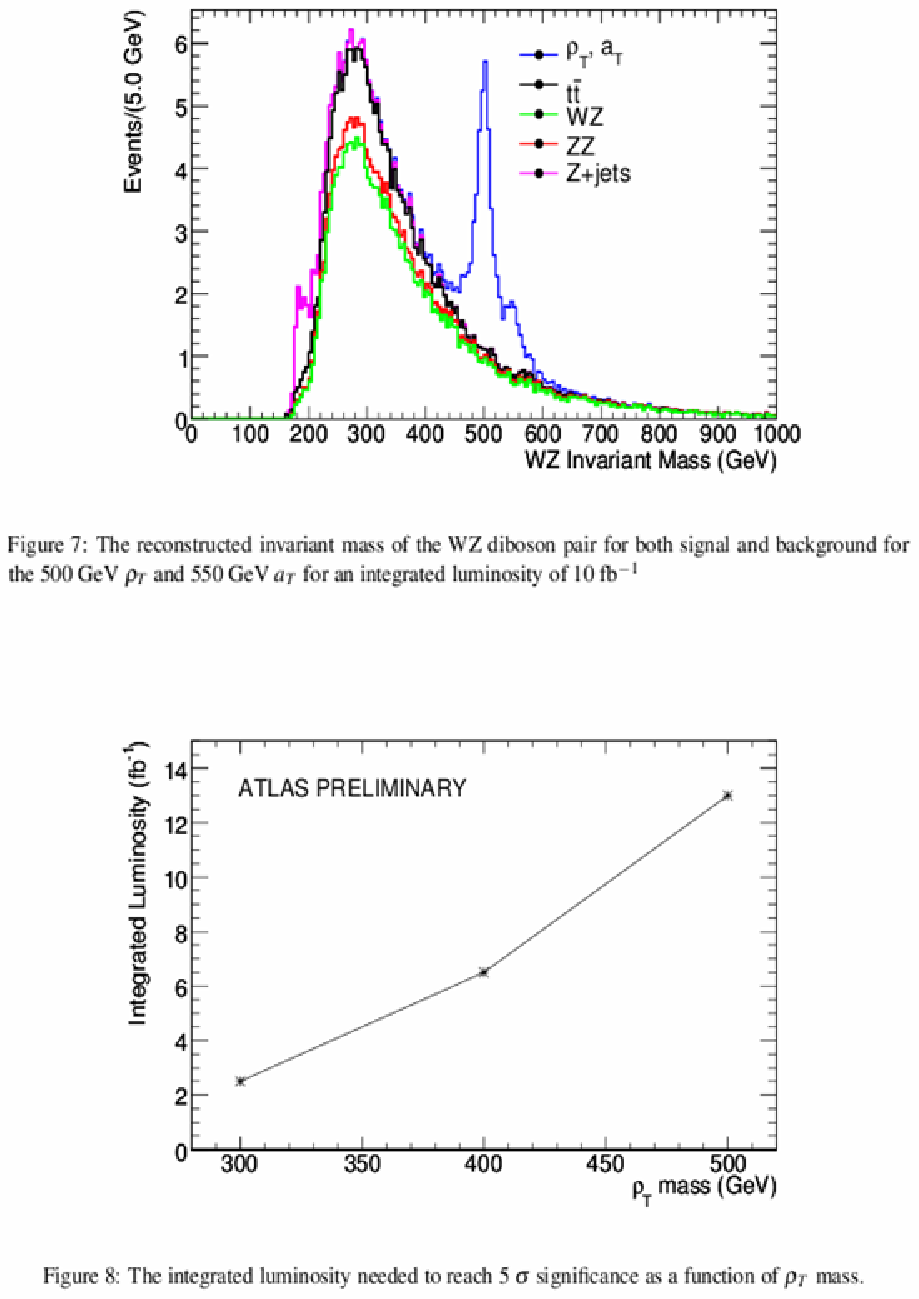}}
    \resizebox{0.49\textwidth}{8.90cm}{\includegraphics{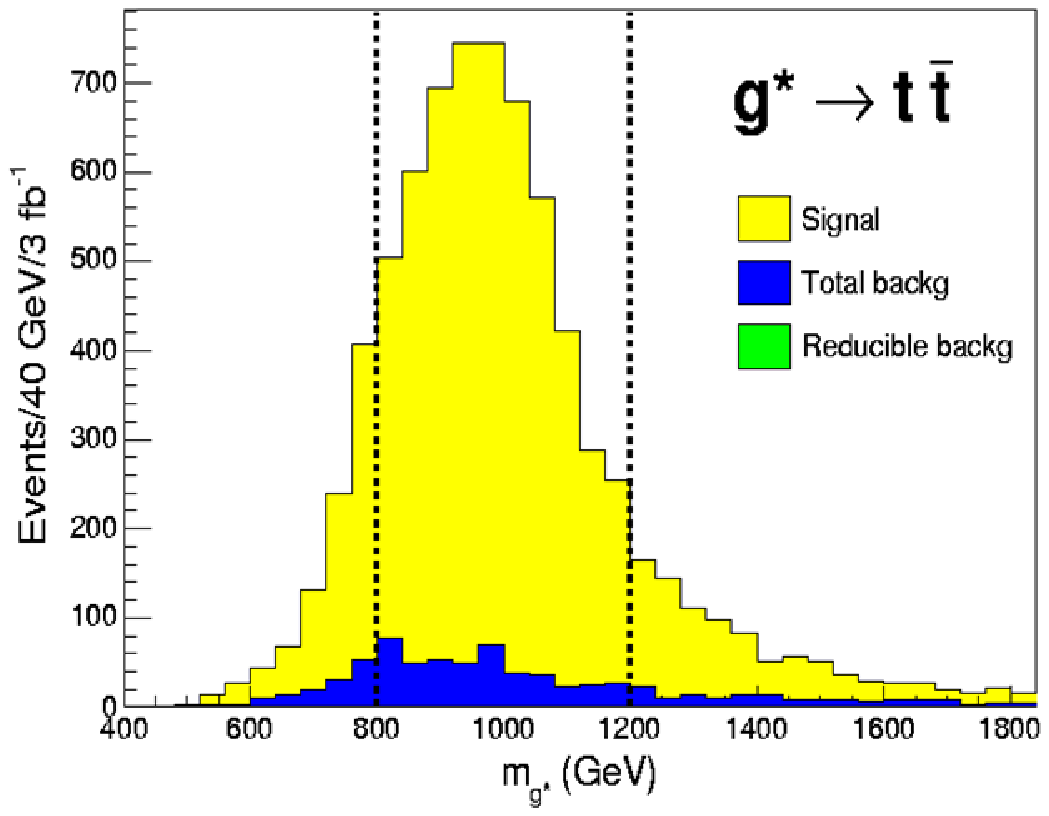}}
  \end{tabular}
\vspace*{-3pt}
\caption{ATLAS studies: Left (7a): technicolor reconstruction and reach.
Right (7b): top-top resonances.}
\label{fig:Fig7}
\end{figure}

The three observed leptons together with the missing transverse energy and the W
invariant mass constraint, a technique developed at the Tevatron, can be used
successfully to reconstruct the full invariant mass of the final state, as shown
in Figure~\ref{fig:Fig7}(left) for ATLAS. Even close-by resonances can be resolved.

\subsection{Two Leptons and Two Jets}

This channel is ideal search field for pair-produced leptoquarks (LQ): hypothetical
particles carrying both lepton and baryon numbers and decaying to a quark and
lepton. Typical simplifying assumptions are that they couple to only one
generation of quarks and leptons and that the interactions are chiral.
Current limits from LEP2 and the Tevatron are $\sim$~250--540 GeV depending on the LQ type.
An update by ATLAS produces a 95 \% CL exclusion below 300 GeV with integrated
luminosity of only 2.8 pb$^{-1}$ and below 800 GeV with 220 pb$^{-1}$, so this is
clearly an early discovery channel.

\subsection{Two Tops}

Even a complex object like a resonance decaying to top-antitop quark pair is no
barrier for the LHC detectors. The mass reconstruction when one W decays to hadrons,
the second to electron or muon plus neutrino is shown in Figure~\ref{fig:Fig7}(right).
Such resonances can arise e.g. in TeV$^{-1}$ Kaluza-Klein extra dimensions models.
The sensitivity will ultimately reach 3.3 TeV.

\subsection{Contact Interactions: Two Leptons or Jets Revisited}

So far we have basically described ``bump-hunt'' type of searches.
\begin{figure*}[!Hhtb]
\centering
        \includegraphics[width=0.90\textwidth,height=8.0cm]{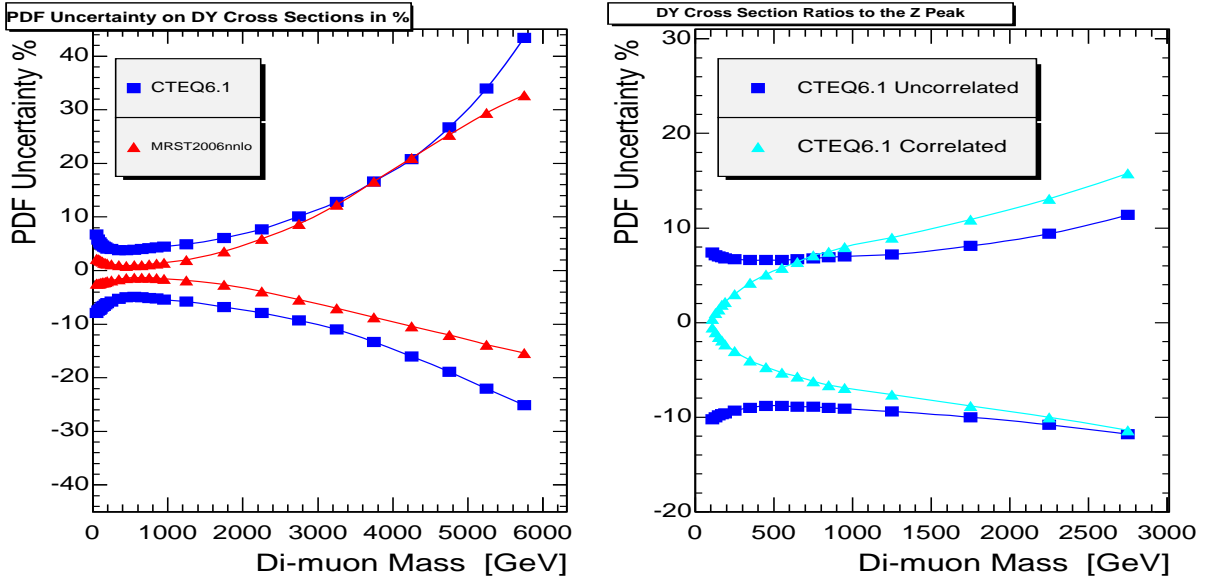}
\caption{CMS PDF uncertainty study for Drell-Yan: Left (8a): uncertainties
on the cross section in the CMS acceptance region for CTEQ6.1 and MRST2006nnlo as
function of mass.
Right (8b): the PDF uncertainties are reduced when using ratios to the Z peak
when correlations are taken into account.}
\label{fig:Fig8}
\end{figure*}
A small digression
are searches where we do not expect to see a resonance, but rather a gradual
deviation from the SM expectation as function of invariant mass. In this subsection we
will limit ourselves to di-leptons and di-jets. Clearly, in contrast to before where
a not-too-wide bump provides an easy handle on the backgrounds e.g. from the side bands
or ad-hoc fits of the background shape excluding the peak region, here a better
understanding of the backgrounds is crucial.
An important component is the uncertainty
on the SM predictions originating from the PDF uncertainties. The modern library
of PDFs LHAPDF provides convenient tools to estimate the size of these effects, as
shown in Figure~\ref{fig:Fig8}. Clearly a normalization to a standard candle like
the Z peak helps to reduce many systematic effects, PDF uncertainties included.

Contact interactions offer a general approach for new interactions at a scale
above the accessible center-of-mass energy. New phenomena can be observed
through virtual effects and interference with the SM amplitudes. The constraints
obtained this way are on the ratio of coupling divided by the scale. A very
popular model searched for today is the ADD scenario for extra dimensions.

\begin{figure}[!Hhtb]
  \centering
  \begin{tabular}{ccc}
    \resizebox{0.33\textwidth}{7.2cm}{\includegraphics{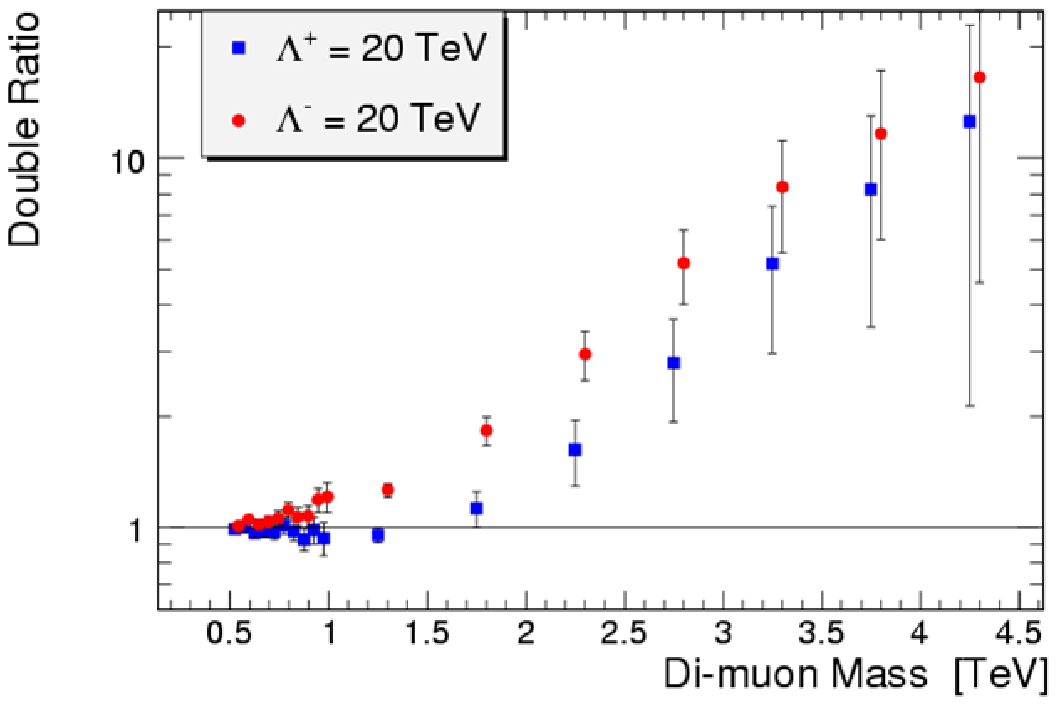}}
    \resizebox{0.33\textwidth}{7.2cm}{\includegraphics{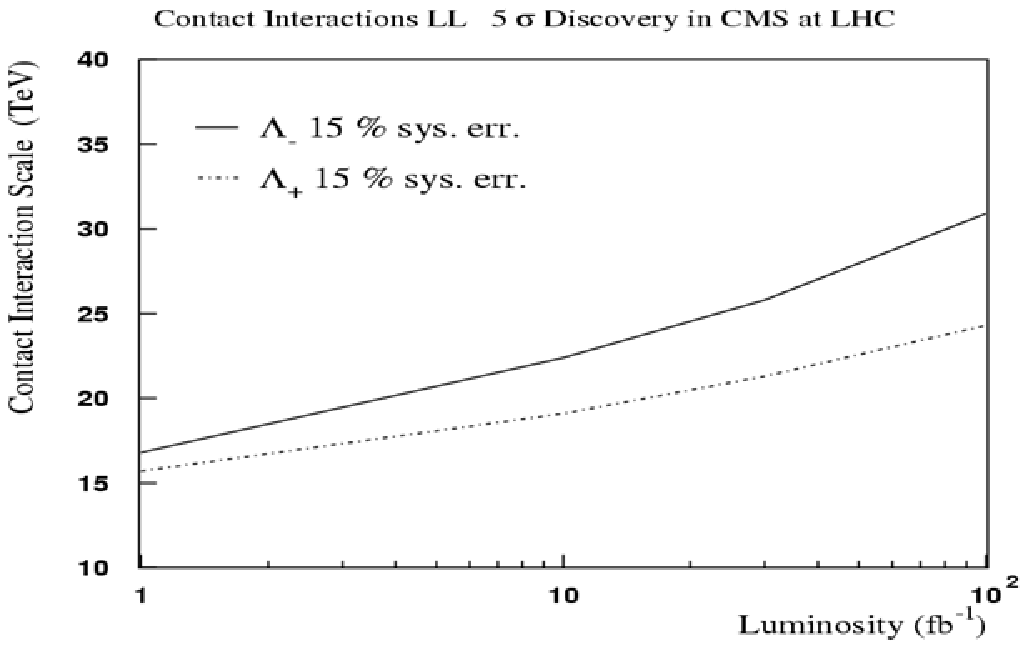}}
    \resizebox{0.33\textwidth}{7.2cm}{\includegraphics{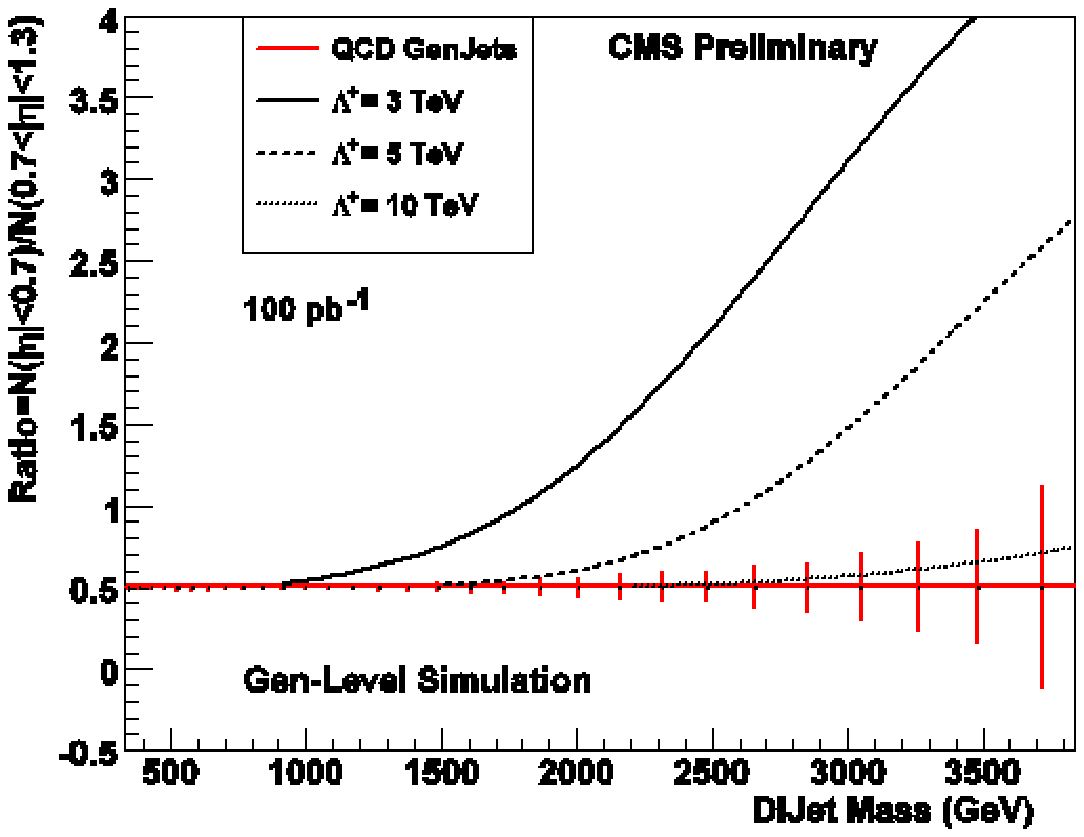}}
  \end{tabular}
\vspace*{-3pt}
\caption{CMS compositeness studies using ratios:
 Left (9a): di-muon study using double ratios of normalized data and Monte Carlo.
If the SM is valid, the ratio is 1 independent of mass.
Middle(9b): di-muon discovery reach.
Right (9c): di-jet study using central/forward ratios.}
\label{fig:Fig9}
\end{figure}

The most promising approach is to use ratios of data from signal enriched and
depleted regions or double ratios of data and Monte Carlo, both normalized
to a well described by the SM region, typically at lower masses, as illustrated
by CMS studies, Figure~\ref{fig:Fig9}. Scales above 15 TeV can be probed already with
1 fb$^{-1}$ in the di-muon channel.

\subsection{``Slow'' Particles}

\begin{figure}[!Hhtb]
  \centering
  \begin{tabular}{ccc}
    \resizebox{0.33\textwidth}{7.2cm}{\includegraphics{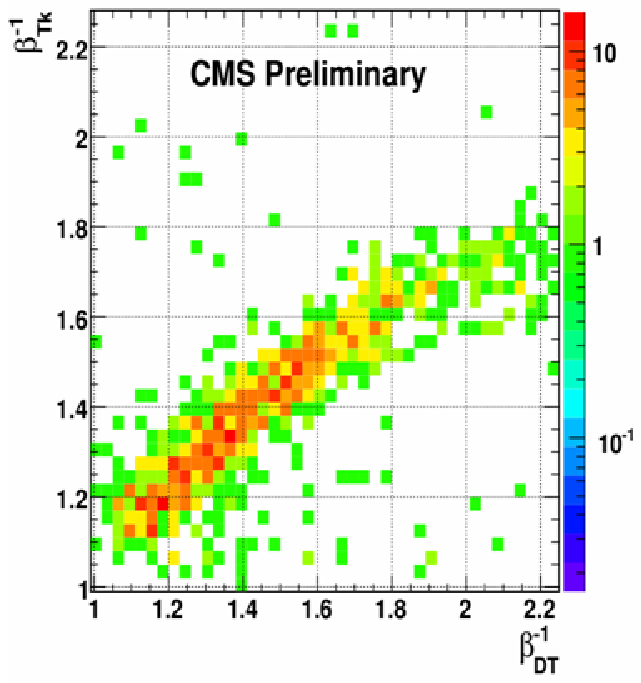}}
    \resizebox{0.33\textwidth}{7.2cm}{\includegraphics{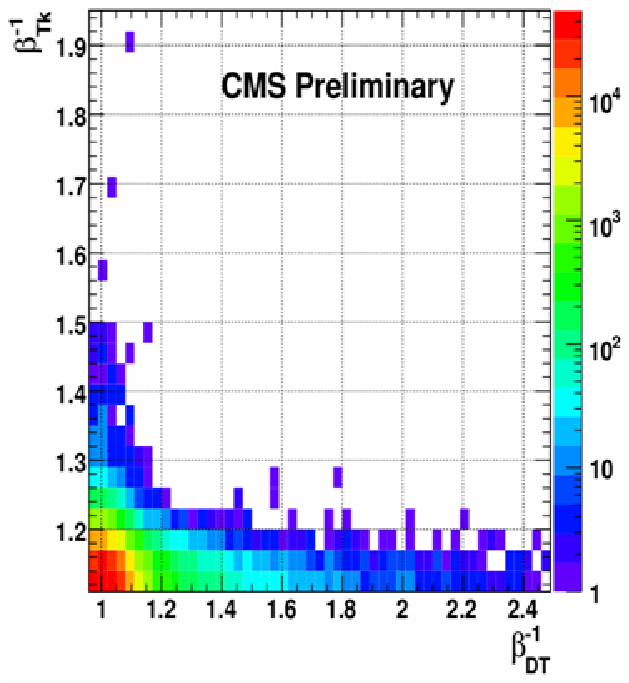}}
    \resizebox{0.33\textwidth}{7.2cm}{\includegraphics{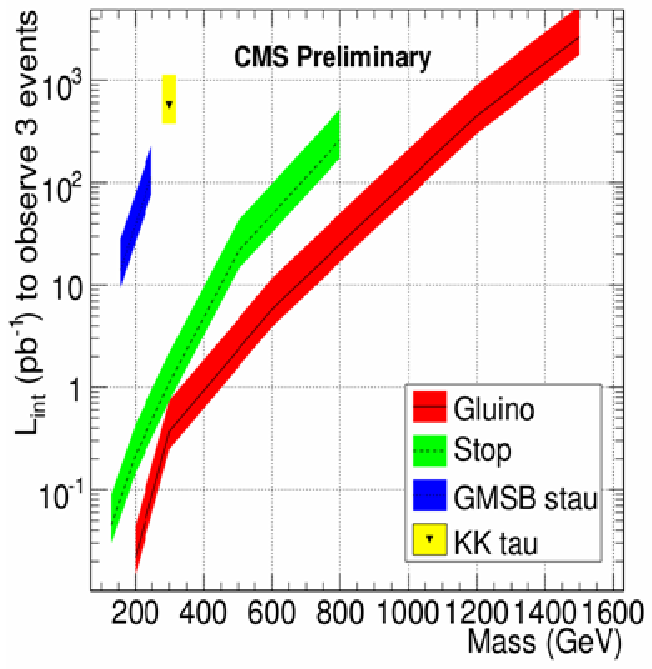}}
  \end{tabular}
\vspace*{-3pt}
\caption{CMS heavy stable charged particles (HSCP) study.
Two measurements of $\beta^{-1}$ in the tracker and the muon drift tubes:
 Left (10a) - signal, 
Middle(10b) - background.
Right (10c): discovery reach.}
\label{fig:Fig10}
\end{figure}

Most particles traverse the LHC detectors at speeds very close to the speed of light.
But if a new particle is very heavy  (hundreds of GeV) it can be slow ($v\ <\ c$) and
hard with sizable momentum above 100 GeV. In this case it is possible to determine
directly the mass of the particle if we can measure both the momentum and $\beta$.
E.g. in CMS $\beta$ can be measured two times independently: using time-of-flight
in the Drift Tube muon stations or specific ionization dE/dx in the tracker. This gives
a powerful handle to extract the signal, making an early discovery possible, Figure~\ref{fig:Fig10}.

\subsection{Unusual Event Shapes}

\begin{figure}[!Hhtb]
  \centering
  \begin{tabular}{cc}
    \resizebox{0.49\textwidth}{8.0cm}{\includegraphics{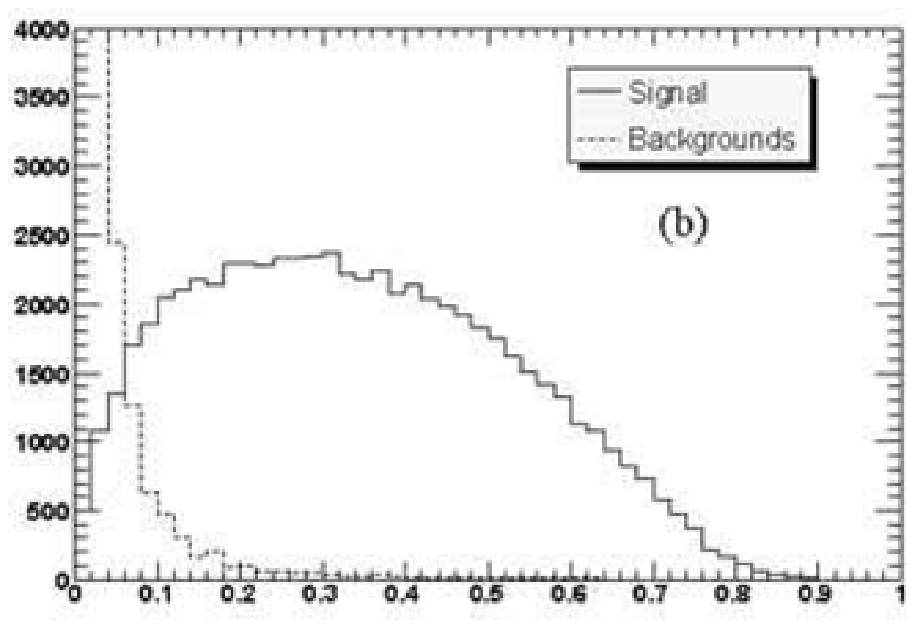}}
    \resizebox{0.49\textwidth}{8.0cm}{\includegraphics{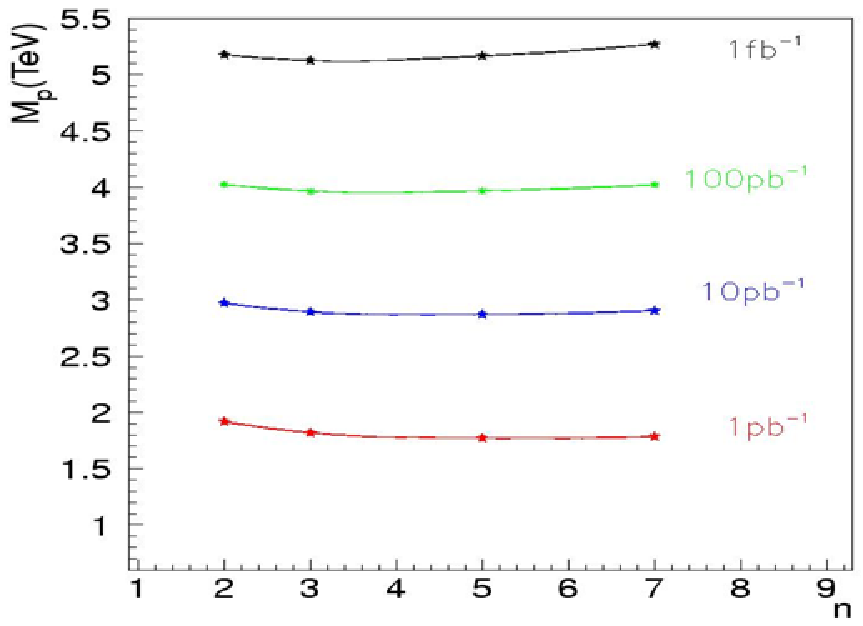}}
  \end{tabular}
\vspace*{-3pt}
\caption{Black hole studies: Left (11a): The striking difference in sphericity (CMS).
Right (11b): ATLAS discovery reach for different luminosities as function of the
number of extra dimensions.}
\label{fig:Fig11}
\end{figure}

So far we have looked at the best measured objects in collider experiments like
leptons, photons, jets and missing transverse energy, which have fuelled many
discoveries over the last decades. Given the large jump in energy, a closer look
at the general event shape and charged or neutral particle multiplicities is
certainly among the first topics and papers to come from the LHC. Besides
testing the SM, the Monte Carlo generator tuning and the extrapolations from the
Tevatron, new phenomena like TeV strings or mini blackholes can be produced
copiously in extra dimensions models with a low enough Planck scale. The
experimental signatures are striking. Mini blackholes for example
will evaporate ``democratically'' through Hawking radiation producing
high multiplicity ``circular'' events, much more spherical than the usual
``jetty'' events, as seen in Figure~\ref{fig:Fig11}.
This could open the prospect of studying quantum gravity at colliders.
Scales up to 5 TeV can be probed with luminosity not even reaching 1 fb$^{-1}$.

\section{OUTLOOK}

The ATLAS and CMS collaborations are about to start exploring the Terascale - the culmination
of long preparations and many thousands of man--years of hard work. This will
open a reach search field for early discoveries: resonances first, hopefully enabling
us to fix the scale for new physics. Many other searches like single photons,
heavy Majorana neutrinos and right-handed bosons, little Higgs, doubly-charged
scalars, isosinglet quarks, same sign top, WW scattering, etc. are not covered in the 
limited time available.

The experiments are concentrating their efforts
to be ready when first collisions take place this fall: detectors, data acquisition systems and
software are in place awaiting first LHC data, A lot of inspired work lies ahead to
understand detectors of this scale and to avoid ``discovering'' detector features.
The years ahead will be very exciting and the prospects are excellent for paper titles
beginning with ``Observation of'' and not with ``Search for''. And be prepared for the
unexpected.

% If you have acknowledgments, this puts in the proper section head.
\begin{acknowledgments}
The author wishes to thank the physicists in ATLAS and CMS for kindly
providing their results, Boaz Klima, Sarah Eno, Albert de Roeck and Karl Jakobs
for their input in preparing the talk.

Last but not least, the author thanks the organizers for the invitation and the kind
hospitality.
\end{acknowledgments}

\end{document}